# Generative Deep Learning for Virtuosic Classical Music

Generative Adversarial Networks as Renowned Composers

Daniel Szelogowski

UW - Whitewater
Computer Science Master's Program
Whitewater, WI
szelogowdj19@uww.edu

**Abstract.** *Current AI-generated music lacks fundamental principles of good compositional techniques. By narrowing down implementation issues both programmatically and musically, we can create a better understanding of what parameters are necessary for a generated composition nearly indistinguishable from that of a master composer.*

## 1. INTRODUCTION

Generative artificial intelligence systems are becoming increasingly popular in everyday usage since the invention of **Generative Adversarial Networks (GANs)** by Ian Goodfellow in 2014 [5]. GANs are specialized neural networks (a type of machine learning algorithm that mimics the way the human brain learns to recognize relationships in data) that learn to choose samples from a special distribution by setting up a competition [12]. This type of **deep learning** model is broken up into two forms:
- A **Generative** model (based on supervised/predictive learning) which describes how a dataset is generated, in terms of a probabilistic model. By sampling from this model, we can generate new data [3]
- A **Discriminative** model (based on unsupervised/non-output learning) which learns to discriminate between data samples from the original dataset and the model generated data [4]

Using this dual-model system, the neural network is able to drive a competition where both players (the Generator and Discriminator) become better over time, creating better-predicted output.

These types of models have grown increasingly popular since their creation and have rapidly developed within the last two years, now being used to generate realistic photography of people (especially faces) and landscapes, handwriting, visual arts, speech/text transcription, language translation, and more recently music. However, music is much more difficult to process for a model due to the number of variables a piece of music can have, such as the notes (including silence), rhythms, dynamics, voicing, phrasing, textures, compositional styles and genres, and so on. Much of current work with music-based GANs lacks musicality or is too error-prone to be performed without the necessity of an external editor manually correcting poorly written phrases. This paper seeks to discuss these issues of current systems and provide a possible solution along with insight into further developing the field to allow for more mature generative compositions.

## 2. RELATED WORK

While there is a growing number of implementations and an increase in interest for music generating GANs, current systems still result in music that is either highly distinguishable from its intended imitation or lacks musicality altogether.

## 2.1 David Foster

Foster's initial model is a **Recurrent Neural Network (RNN)** which utilizes a **Long Short-Term Memory (LSTM)** architecture to generate music through **MIDI** sequences [2]. RNNs feature nodes connected together as a directed graph, wherein their internal state (memory) allows them to process input sequences of variable length [13]. Using LSTM, the neural network's weights are no longer hindered by speed as we go further right in the network, allowing the analysis of entire sequences of data rather than just single data points — this will form the basis of our own model. The model uses this two-stack LSTM system to form an attention mechanism, allowing the model to decode information encoded in previous layers to better predict a sequence's continuation [2].

Foster's model was trained on all six of Johann Sebastian Bach's Cello Suites — while this is a small set of data which could have the potential to work well, the six suites are broken down into six movements of varying themes, tonalities, and rhythmic variation; as such, only very short sequences are able to be generated to sound well, but generally replicate existing sequences rather than generating completely original ones.

## 2.2 MuseGAN

While Foster discusses the RNN model briefly [2], the MuseGAN model sees much more focus; this model is similar to Foster's implementation but allows for the input of multiple noise vectors to perform training and generation of polyphonic (multiple-voice) compositions [14]. With a MuseGAN, one can control the amount of noise for parameters such as melodic generation, rhythmic groove, chord/harmonic generation, and the generated style [2].

## 2.3 Google Doodle

In 2019, the Google Doodle team worked with Google's PAIR (People + AI Research) and Magenta (Generative Music and Art Research) teams to create a web application inspired by Johann Sebastian Bach, whose birthday the application celebrated [6]. The application allows users to write a simple melody line on a musical staff, to which the generative model would compose three harmony voices in Bach's contrapuntal "Chorale" style.

While the model does mimic the four-voice texture of a typical Bach chorale, the application itself has many limitations, such as the inability to modify the key signature or meter (or time signature) of the piece or write note durations longer or shorter than 4th and 8th notes, though the user may choose to leave some note "spaces" empty. As well, the application features many musical issues:
- The composition generated by the model lacks clear tonality even given a very melodic line, such as one taken from an actual chorale
- Accidentals, or flattened/sharpened notes outside of the key signature, are sporadically thrown in with no melodic or harmonic reason
- While the voices are meant to move independently, their relationship is important; the model has no clear concept of one line intruding upon another, or voice-crossing, causing what is generally an intentional effect to become accidental and disturb the harmony of the piece

Many of these issues are due to the simplicity of the implementation, but also with the model itself. The developers trained the model on 306 Bach chorales (of the ~419 total) — while this may be effective for compositions that are tonally and rhythmically similar, this is a highly ineffective approach due to the extreme amount of tonal and melodic variance within the Bach chorales.

## 2.4 Huawei

As a demonstration of their newest smartphone in February of 2019 (the Mate 20 Pro), consumer electronics manufacturer Huawei created a model to be run as an Android application which would listen to and train on Franz Schubert's Symphony No. 8, a piece famously left unfinished due to only having two of the standard four movements of a typical symphony [7].

Having earned the popular title of the "Unfinished Symphony," Huawei sought to use the trained model to generate the third and fourth movements of the piece before having the newly completed piece performed by a live orchestra. Two major issues stand out from this approach:
- The model was only trained on half of a symphony — typically all four movements have varying musical styles, causing a lack of variation in the generated movements, as well as lacking the compositional techniques that would have been beneficially trained from Schubert's other symphonies
- The final output generated by the model was manually edited by the orchestra director to remove errors in musicality and create a better sense of phrasing

As a result, the generated movements were musically similar to the first two, overly repetitive, and lacked independence from an external editor.

**2.5 OpenAI Jukebox**
Jukebox is a generative neural network with a language model created in late April of 2020 that has been used primarily in studying generative popular music such as classic rock and pop by letting the model finish parts of well-known songs, including voice and instruments [8]. Rather than utilizing two separate input layers to separate the vocals and instruments, however, the model condenses entire audio files into discrete data to create a raw audio model, rather than OpenAI's previous project MuseNET which trained on MIDI data similar to Foster's RNN model [9].

### 3. APPROACH

In order to narrow down a solution to the aforementioned issues, my implementation will take place in a program with two components: a generative deep learning algorithm and a classification algorithm. Using a GAN along with a **Support Vector Machine (SVM)**, an algorithm that analyzes data to perform binary classification (as one of two categories within a set) with as large of a margin as possible between the two classes/categories. Through regression, the SVM builds a strong sense of discrimination between two classes and has great potential to distinguish data as one of them or the other.

Together, the two algorithms will perform the following tasks:
- GAN: Read in each MIDI file and learn from the data
- GAN: Attempt to generate a new composition of similar style
- SVM: Classify the data in binary - by composer or not by composer
- GAN: Strive to increase confusion for SVM

The GAN will learn from evolution to generate pieces more similar to the composer and make it more difficult to differentiate between real and generated. Ideally, the GAN should be able to recreate or replicate virtuosic compositional techniques, and we should strive for the SVM to have less than 50% accuracy as the network becomes better at composing.

### 4. IMPLEMENTATION

My algorithm follows that of Foster's "RNN With Attention" model, utilizing a two-stack LSTM system to allow attention dedication to musical pitch and another to note duration; the Python application itself employs the **Keras** deep learning API which is built upon Google's **Tensorflow** library, as well as the libraries **music21** for extracting MIDI data, **Pickle** to create a data store for the pitch and duration sequences, and **Sci-kit Learn** for the SVM algorithm. The two input sequences (or the previous layer's hidden state, depending on the recurrence) will feed into the recurrent layer one step at a time, continually updating its hidden state. Once this state is finalized, it will form a vector the same length as the sequence length we train it on, which can then be fed into a **Dense** layer. This layer will have

**softmax** output (referring to the "softargmax" function) to predict a distribution for the sequence's next note. A full diagram of this model can be seen in Figure 1 (all implementations can be found in [15]).

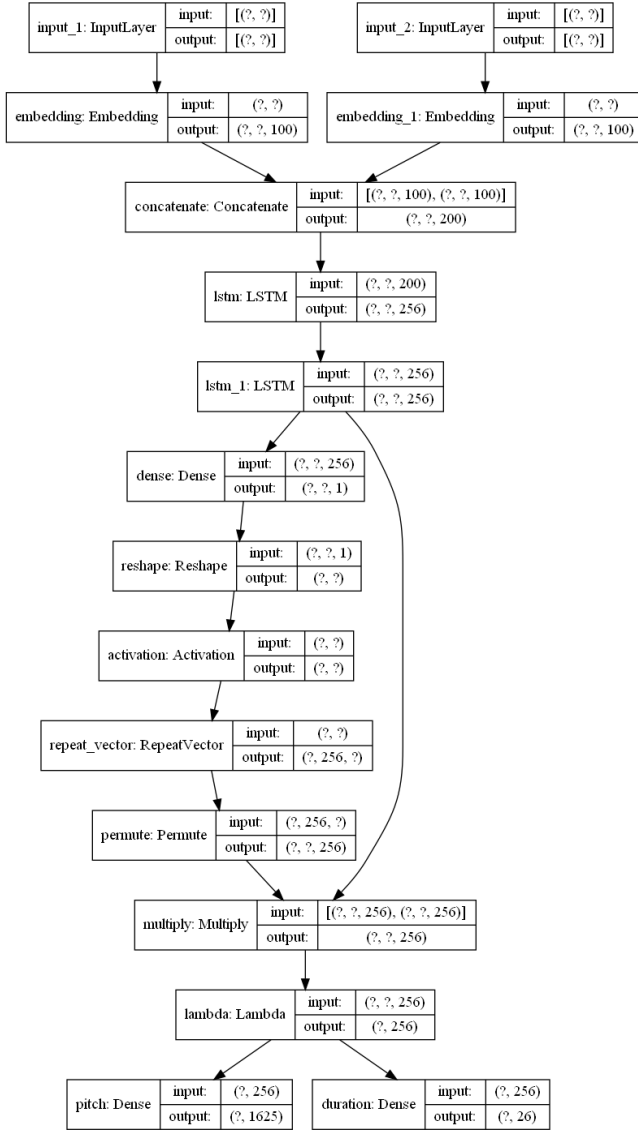

**Figure 1: The RNN model after training on all 24 caprices**

After training on each composition to a user-defined number of **epochs** (or evolutions), the model is ready to compose. During the early evolutions, the model will stick to writing one note repeatedly with the same rhythm, eventually becoming more confident over time to generate small atonal note runs and eventually full sequential passages with interesting rhythmic durations centered firmly within a set musical key. Figure 2 and Figure 3 show the result of a model that has been trained thoroughly (250+ epochs) and can produce genuine musical phrases.

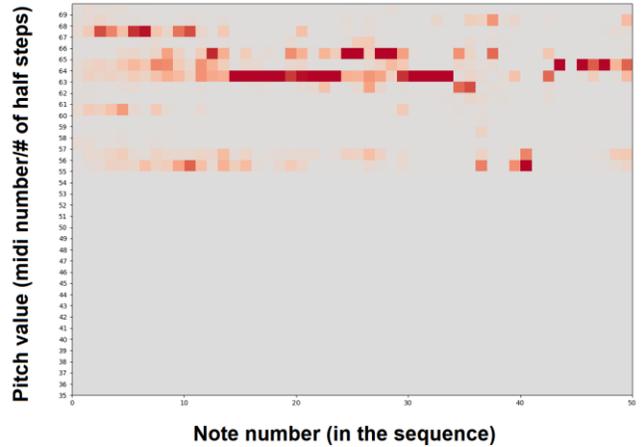

**Figure 2: Distribution of notes over time — darker squares represent a stronger certainty of being the next pitch**

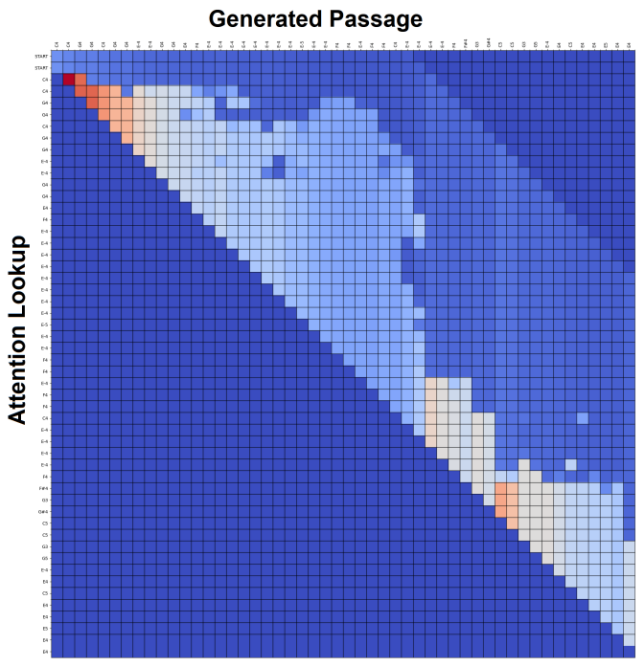

**Figure 3: A heatmap of the model's attention on the recurrent layer's hidden state corresponding to a particular note in the sequence over time — a "hotter" square denotes more attention given**

## 5. EVALUATION

The SVM classifier is the third and final step in the GAN's compositional journey. To classify the compositions, the MIDI files must first be converted into a numeric dataset. To achieve this, we can apply the following steps:

1. Read in all notes and durations from the data file
2. Create an array of each note's pitch space value (or the number of half steps/midi value, an integer) with its duration (in fraction form) added together in the form *Note.Duration*. For any instances of a chord or double stop (where more than one note is played at once), we can multiply the notes of the chord together to create a unique identity that is less collision-prone than through addition or XOR
3. Write each note in the array as a column in a CSV table file. Each piece will be an additional row added to this table

Utilizing this will help to reduce overhead for performing multiple classifications, as the data for the composer's pieces will already be written to a file and only the GAN's composition will need to be converted. Afterward, the data can be compiled into one large N-Dimensional array containing the data of all of the composer's pieces along with one composition by the GAN, and the SVM can run its predictions. Given that SVM is such a strong discriminator, the fairest kernel to apply is linear — all others will either always be correct or overshoot.

To break down the training into multiple portions and discover solutions to many of the issues mentioned in *Section 2*, the model was divided up into four sub-models which were trained to write 32 note sequences at a time (for a range of 400-1200 notes allowed total, the typical range for the caprices) for a total of 60 compositions per model:

- **Paganini** - A model trained on all 24 caprices
- **Melodics** - A model trained on caprices 20-24 due to their similar melodic structure, phrasing, and rhythms
- **Moderato** - A model trained on caprices 2-4, 7, 9, 11, 14, 17, 18, 21, 23, and 24 for their similar "moderate" rhythmic pace
- **Presto** - A model trained on caprices 1, 5, 6, 8, 10, 12, 13, 15, 16, 19, 20, and 22 for their similar "rapid" rhythmic pace

After initially training only the Paganini model, I found that the generated compositions were either mostly atonal, quickly switched between slow and rapid tempos, repeated many notes, and generally failed to recover its attention after each sequence was written. To mitigate this, I created the Melodics model which yielded the best results of all four models, since the pieces it was trained on were so similar in key, tempo, phrasing, and rhythmically in general; after which I divided the entire dataset into two halves to form the Moderato and Presto models, both now having twice as many MIDI files to train on as Melodics, but half of Paganini. All models were trained to at least 250 epochs; all four suffered from similar issues as mentioned with the Paganini model, but these were greatly lessened through the data subdivision.

Presto and Moderato produced the most original, truly musical sounding compositions, with Paganini generally being the worst overall. However, regardless of their originality and musicality, the SVM was able to more easily distinguish that these were not by the composer than some that I believed did not sound as musical. For all four models as well, I divided the test into two portions:
- **Full Dataset**
- **Individual Dataset**

The models generally performed better on their own individual datasets than with all 24 pieces, but not so steep as to promote a strong inclination that the models are ineffective against the entire set given the average difference between the two sets being ~2%.

## 6. DISCUSSION

After running the full SVM examination on the four models, I discovered that the models were able to generate multiple pieces which fell below the hypothesized 50% classification rate, with the lowest overall score being 18.29% and the best across both datasets averaging 56%. Three of the most (aurally) realistic musical compositions and two of the most SVM-confusing compositions generated were prepared in a quiz (which can be found in [16]) which contained 15-second snippets of each of the 5 pieces and of 5 randomly chosen caprices by the composer in a shuffled order, with a final question asking the survey taker if they are a music teacher, student, amateur/hobbyist, or listener, the division of which can be seen in Figure 4 distributed among 25 surveyors. The results of this quiz are shown in Figure 5 and Figure 6. Audio recordings of these and other generated pieces can be found in [1].

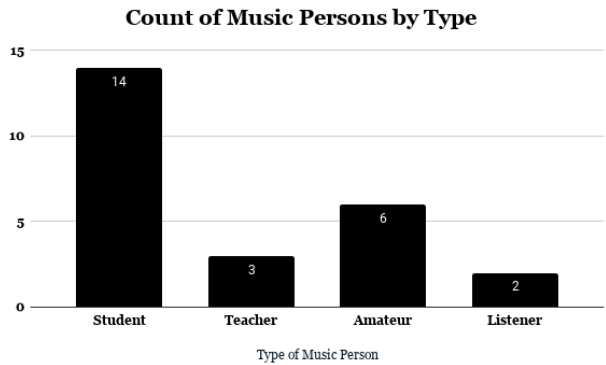

**Figure 4: Distribution of survey takers by musical experience**

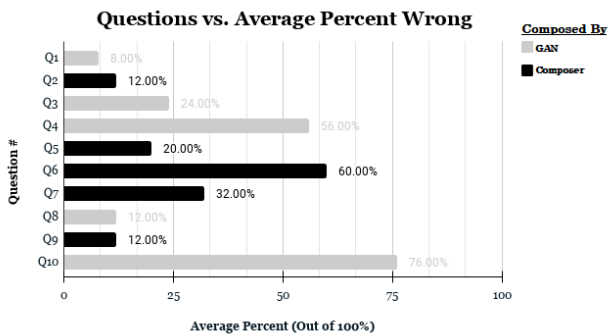

**Figure 5: Questions by composition origin versus average percent wrong**

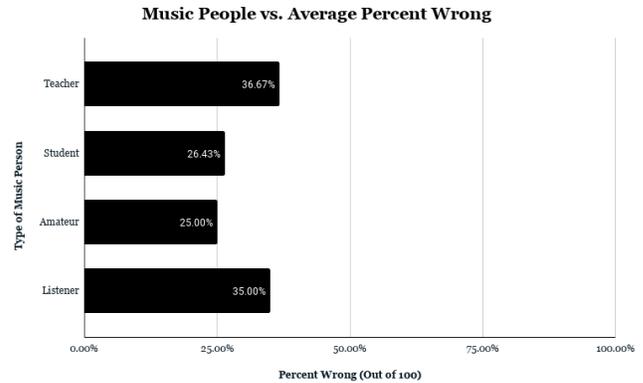

**Figure 6: Survey takers by musical experience versus average percent wrong**

Ideally, the success of this approach to training may be extended into a more mature form of GAN which selectively pairs input data to better discriminate sequences by similarity and train based on musical intention, rather than just by notation (much like the errors mentioned in Foster's original model in *Section 2.1*).

## 7. CONCLUSION

This paper focused on narrowing down the most common issues with current generative music systems, primarily with the system of training these various models, and working toward a solution which would perform well against a large-margin discriminator as well as actual people and remove the need for a human editor for a generative composition to sound musically "good". My experiments revolved around both narrowing and broadening training exercises between concisely similar versus relatively similar compositions, rather than just the entire dataset, in order to discover what the models struggled with the most and why they would hang up on a particular note, sequence, or lose attention entirely and write completely atonally.

In the future, this type of model could be expended into a much more advanced form with a stronger discrimination system and utilize an infinite number of voices/instruments to create a more advanced MuseGAN — ideally a model with the capability to potentially train on full orchestral MIDIs and create more musically genuine third and fourth movements to Schubert's Unfinished Symphony by training on his other 7, both wholly and on the last two movements selectively and infer through strong discrimination which of the two is more realistic for example, or creating Beethoven's 10th Symphony by training on his 9 others. As well, a model could have an additional system of language and vocal prediction, allowing for the realistic generation of solo voice and choral pieces, and potentially full operas through textual training.

## ACKNOWLEDGMENTS

My sincerest gratitude goes to Noah Schaffrick for his contribution of live recordings of many of the best compositions generated by the neural network, as well as to the music faculty of the University of Wisconsin - Whitewater for their interest, enthusiasm, and great contribution in bringing awareness to this research and providing additional surveyors.